\documentclass[journal=nalefd,manuscript=article]{achemso}

\usepackage{chemformula} 
\usepackage[T1]{fontenc} 
\usepackage{amsmath}
\usepackage{gensymb}
\usepackage{siunitx}   
\usepackage{mhchem}    
\usepackage{algpseudocode}
\usepackage{upgreek}
\usepackage{textcomp}

\author{Fabia F. Athena}
\affiliation{Department of Electrical Engineering, Stanford University, CA, USA}
\altaffiliation{These authors contributed equally}
\email{fathena@stanford.edu}

\author{Jonathan Hartanto}
\affiliation{Department of Electrical Engineering, Stanford University, CA, USA}
\alsoaffiliation{Department of Materials Science and Engineering, Stanford University, CA, USA}
\altaffiliation{These authors contributed equally}
\email{jh99@stanford.edu}

\author{Matthias Passlack}
\affiliation{Corporate Research, Taiwan Semiconductor Manufacturing Company, Ltd., San Jose, CA, USA}

\author{\mbox{Jack C. Evans}}
\affiliation{Department of Electrical Engineering, Stanford University, CA, USA}

\author{\mbox{Jimmy Qin}}
\affiliation{Department of Electrical Engineering, Stanford University, CA, USA}

\author{Didem Dede}
\affiliation{Department of Materials Science and Engineering, Stanford University, CA, USA}

\author{Koustav Jana}
\affiliation{Department of Electrical Engineering, Stanford University, CA, USA}

\author{Shuhan Liu}
\affiliation{Department of Electrical Engineering, Stanford University, CA, USA}

\author{\mbox{Tara Pe\~na}}
\affiliation{Department of Electrical Engineering, Stanford University, CA, USA}

\author{Eric Pop}
\affiliation{Department of Electrical Engineering, Stanford University, CA, USA}

\author{Greg Pitner}
\affiliation{Corporate Research, Taiwan Semiconductor Manufacturing Company, Ltd., San Jose, CA, USA}

\author{\mbox{Iuliana P. Radu}}
\affiliation{Corporate Research, Taiwan Semiconductor Manufacturing Company, Ltd., Hsinchu, Taiwan}

\author{\mbox{Paul C. McIntyre}}
\affiliation{Department of Materials Science and Engineering, Stanford University, CA, USA}

\author{\mbox{H.-S. Philip Wong}}
\affiliation{Department of Electrical Engineering, Stanford University, CA, USA}
\email{hspwong@stanford.edu}

\title[An \textsf{achemso} demo]
  {Gate Dielectric Engineering with an Ultrathin Silicon-oxide Interfacial Dipole Layer for\\ Low-Leakage Oxide-Semiconductor Memories}


\begin{document}

\begin{tocentry}

  \includegraphics[width=1\textwidth]{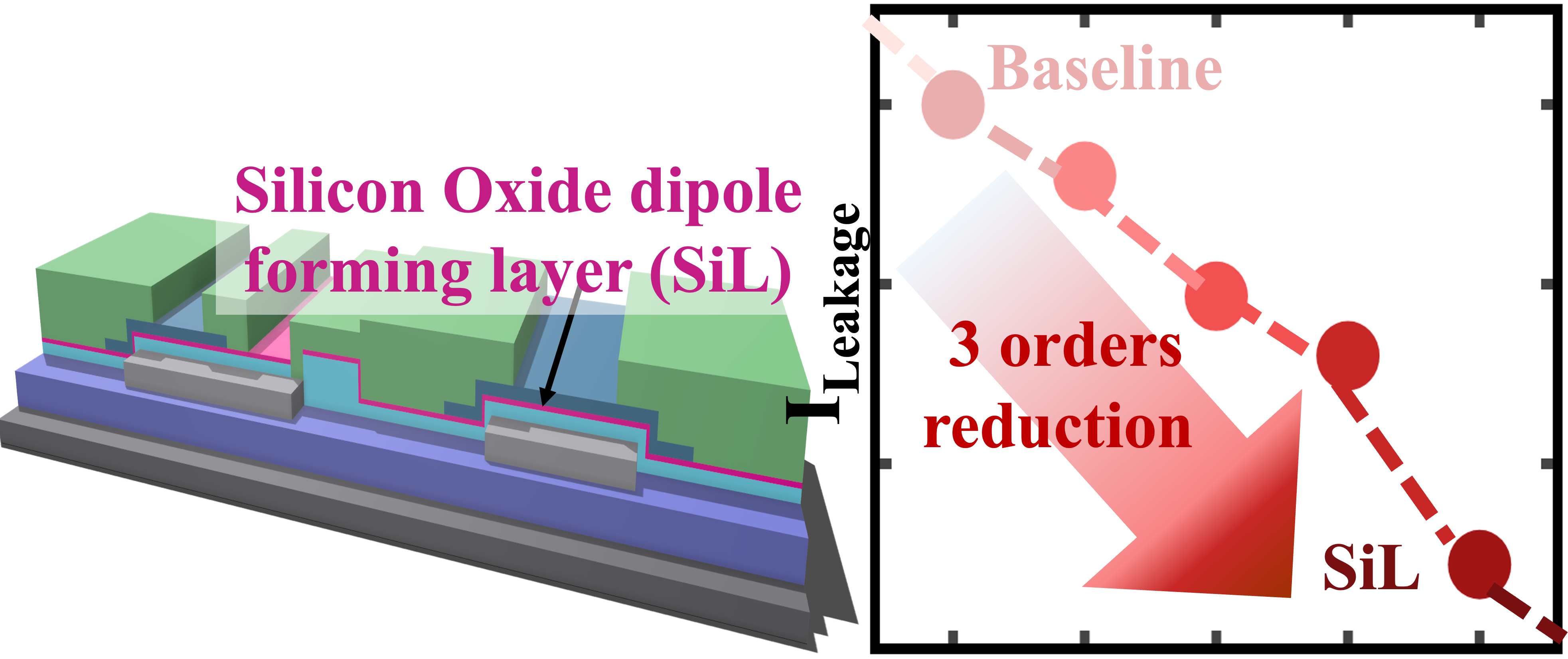}

\end{tocentry}

\begin{abstract}
We demonstrate a gate dielectric engineering approach leveraging an ultrathin, atomic layer deposited (ALD) silicon oxide interfacial layer (SiL) between the amorphous oxide semiconductor (AOS) channel and the high-k gate dielectric. SiL positively shifts the threshold voltage (V\textsubscript{T}) of AOS transistors, providing at least four distinct V\textsubscript{T} levels with a maximum increase of 500 mV. It achieves stable V\textsubscript{T} control without significantly degrading critical device parameters such as mobility, on-state current, all while keeping the process temperature below 225 \textdegree C and requiring no additional heat treatment to activate the dipole. Positive-bias temperature instability tests at 85 \textdegree C indicate a significant reduction in negative V\textsubscript{T} shifts for SiL-integrated devices, highlighting enhanced reliability. Incorporating this SiL gate stack into two-transistor gain-cell (GC) memory maintains a more stable storage node voltage (V\textsubscript{SN}) (reduces V\textsubscript{SN} drop by 67\%), by limiting unwanted charge losses. SiL-engineered GCs also reach retention times up to 10,000 s at room temperature and reduce standby leakage current by three orders of magnitude relative to baseline device, substantially lowering refresh energy consumption.
\end{abstract}

\section{Introduction}

The persistent performance gap between processor speeds and memory bandwidth, commonly referred to as the "memory wall",\cite{aabrar2024reliability,yoo2024atomic} underscores the need for scalable, high-density, high-performance three dimensional (3D) chips enabled by back-end-of-line (BEOL) memory and logic.  Amorphous oxide-semiconductor field-effect transistors (AOSFETs) are attractive for such BEOL integration due to compatibility with BEOL processes, low-temperature deposition, high-density integration potential, and excellent thickness uniformity achievable via atomic layer deposition (ALD) techniques \cite{mukai2002proposal,choi2024c,fortunato2012oxide,liu2024first,yoo2024atomic,10504765,10873499,jana2025key,ryu2024high}.Across envisioned uses—including 3D or 4F² (F is minimum feature size) dynamic random access memory (DRAM), ferroelectric field effect transistor (FeFET) based memory devices, power-delivery switches, and exploratory back-end-of-line (BEOL) logic—precise threshold-voltage (V\textsubscript{T}) control is indispensable\cite{atsumi2012dram,chen2024novel,luo2021emerging,fujii2024oxide,hur2024oxide,ye2025high}. For gain-cell (GC) memories\cite{shukuri2002semi,liu2024first,yoshida2024comparative,bonetti2020gain} in particular, V\textsubscript{T} directly sets retention time and write margins. An excessively negative V\textsubscript{T} increases subthreshold leakage and decreases retention time, whereas an overly positive V\textsubscript{T} within a fixed operating voltage hampers charge transfer during write operations, making writes unreliable. Elevated temperatures exacerbate the V\textsubscript{T} shift because additional thermal energy increases the concentration of oxygen vacancies and free carriers in the OS, driving V\textsubscript{T} further negative\cite{si2021high,conley2010instabilities,evans2025optimization,shen2024reliability,aabrar2024reliability}. Achieving control over V\textsubscript{T} is therefore essential. Conventional V\textsubscript{T}-tuning techniques such as doping, annealing, or passivation often introduce undesirable side effects, such as reduced field-effect mobility ($\upmu_{\text{FE}}$), degraded subthreshold slope (SS), and reliability issues linked to hydrogen release and bias instability\cite{si2021high,wahid2023effect,yoo2024atomic,kita2009origin,kita2010formation,nakata2013influence,lee2025high}. In silicon MOSFETs, interface‑dipole engineering shifts V\textsubscript{T} positively or negatively via “p‑type” or “n‑type” dipoles by inserting ultrathin layers such as AlO\textsubscript{x} or LaO\textsubscript{x} into the gate stack, which sets an interfacial potential step that moves the channel band edges relative to the gate \cite{HUANG2025114073,9667116,sivasubramani2007dipole,athena2025orthogonal,lin2011atomic,cao2023density}. This offers a promising alternative approach for V\textsubscript{T} control in two dimensional (2D) FETs\cite{jskNL,jskVLSI}. In previous works\cite{athena2025orthogonal}, we showed inserting AlO\textsubscript{x} into the gate dielectric effectively shifts the V\textsubscript{T}  in AOS transistors.

In this work, we set out to investigate what other oxides could be used to tune V\textsubscript{T}  in the positive direction for AOS devices, to shed light on the use of a possible broader set of materials. This is also necessary to achieve volume-less V\textsubscript{T}  tuning for AOS gate-all-around (GAA) FETs. We study interface dipole engineering as a volume-less V\textsubscript{T} tuning technique for amorphous indium-tungsten oxide (IWO) AOSFETs by integrating an ultrathin silicon oxide (SiO\textsubscript{x}) interfacial layer. Ultra-thin SiO\textsubscript{x} layers are deposited via plasma-enhanced atomic-layer deposition (PEALD) at 200 \textdegree C without any additional annealing treatment. The SiL-engineered AOSFETs achieves positive shifts in V\textsubscript{T} up to approximately 500 mV without significantly compromising device performance metrics, such as field effect mobility ($\upmu_{\text{FE}}$) or subthreshold slope (SS). Additionally, we evaluate the impact of the SiL on positive bias temperature instability (PBTI) at elevated temperature and demonstrate its effectiveness in mitigating negative V\textsubscript{T} shifts under extended positive bias stress conditions. As an example of the application of this V\textsubscript{T}-tuning technique, we integrated the SiL into a two-transistor (2T) GC memory. Setting V\textsubscript{T} higher significantly reduces the V\textsubscript{SN} drop and substantially lowers leakage current, thereby improving retention and reducing refresh-energy requirements.

\section{Results and Discussion}

\begin{figure}
  \centering
  \includegraphics[width=1\textwidth]{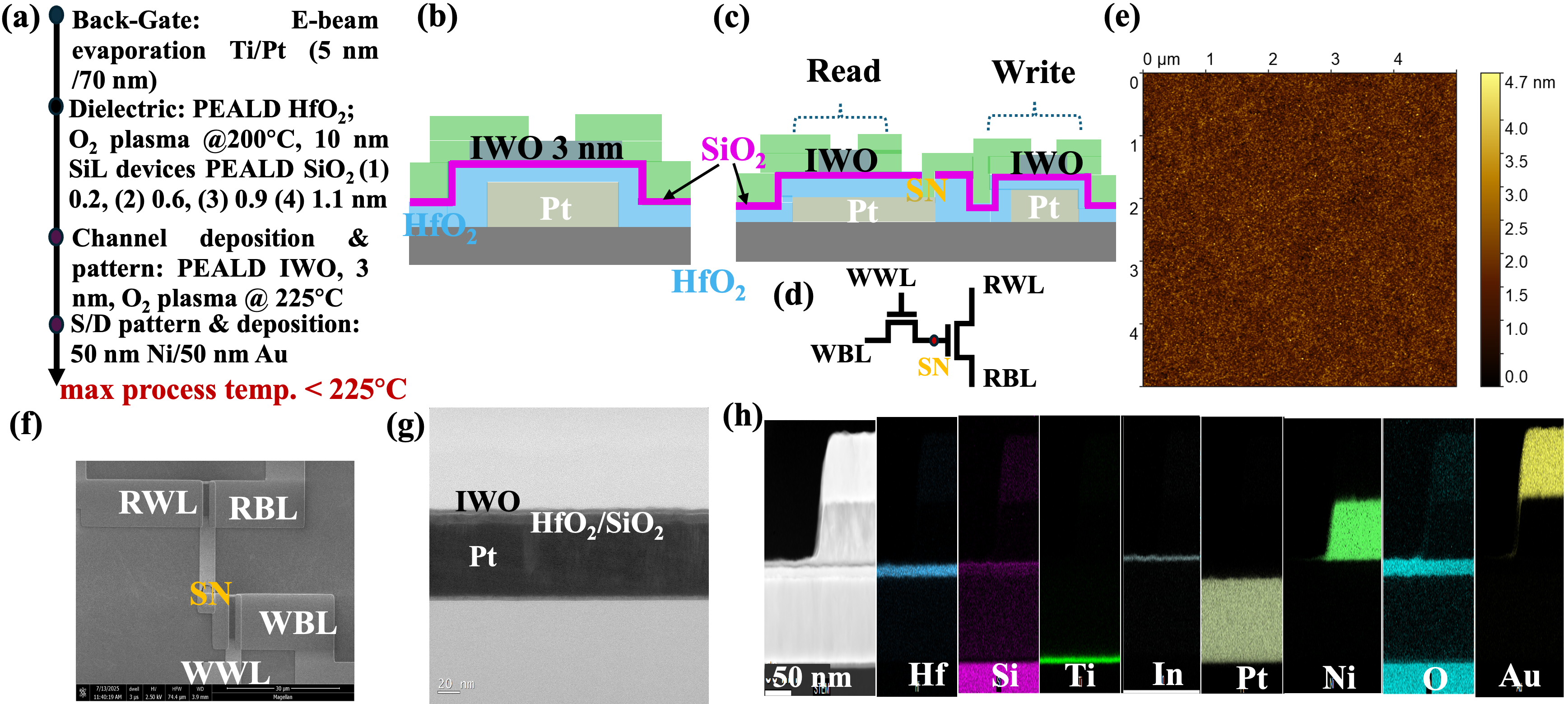}
  \caption{(a) Fabrication process flow of AOSFETs. Schematic illustration of (b) FET structure incorporating a SiO\textsubscript{x} dipole layer, (c) 2T GC, where the source of the WTR is connected to the gate of the RTR. (d) Circuit diagram of the corresponding n-n 2T GC highlighting the SN. (e) Top-view AFM image of IWO channel showing smooth surface of the film, revealing a relatively smooth surface (R\textsubscript{q} $<450 \pm20$ pm). (f) Top view SEM image of the fabricated GC showing RWL, RBL, WWL and WBL. Source of the write transistor is connected to the gate of the read transistor through SN. (g) Cross-sectional TEM image of the FET structure. (h) EDS elemental mapping illustrating uniform distribution of Hf, Si, Ti, In, Pt, Ni, O, and Au across the device cross-section.
}
  \label{fig:1}
\end{figure}

Figure~\ref{fig:1}(a) shows the fabrication process flow of AOSFETs. For SiL-integrated devices, ultrathin SiO\textsubscript{x} interfacial layers with thicknesses of 0.2 nm (SiL1), 0.6 nm (SiL2), and 0.9 nm (SiL3), 1.1 nm (SiL4) are deposited \emph{in situ} with \ce{HfO2}. Details of the fabrication process are provided in the Supporting Information. Figure~\ref{fig:1}(b) schematically illustrates AOSFET structure incorporating the SiO\textsubscript{x} based interfacial dipole layer which lies between the amorphous IWO channel and the high-$k$ \ce{HfO2} gate dielectric. Figure~\ref{fig:1}(c) illustrates a 2T GC structure comprising a write transistor (WTR) and a read transistor (RTR). The source of the WTR is connected to the gate of the RTR forming a storage node (SN). Figure~\ref{fig:1}(d) provides the corresponding circuit diagram of 2T GC. Figure~\ref{fig:1}(e) shows a top‑view atomic force microscope (AFM) image of the IWO channel in the OSFET, revealing a relatively smooth surface (R\textsubscript{q} $<450 \pm20$ pm). Details of the AFM procedure are described in the Supporting Information. A top view scanning electron microscope (SEM) in Figure~\ref{fig:1}(f) shows the layout of RTR, WTR, SN and read word line (RWL), read bit line (RBL), write word line (WWL), and write bit line (WBL). Details of the SEM measurement are provided in the Supporting Information. Figure~\ref{fig:1}(g) presents a cross-sectional transmission electron microscope (TEM) image of the fabricated device structure, which shows well-defined layers of IWO channel, \ce{HfO2}/SiO\textsubscript{x} gate stack and Pt back-gate electrodes. Energy-dispersive spectroscopy (EDS) elemental mapping further confirms the uniform deposition and precise composition of different layers across the device stack as shown in Figure~\ref{fig:1}(h).

\begin{figure}
  \centering
  \includegraphics[width=1\textwidth]{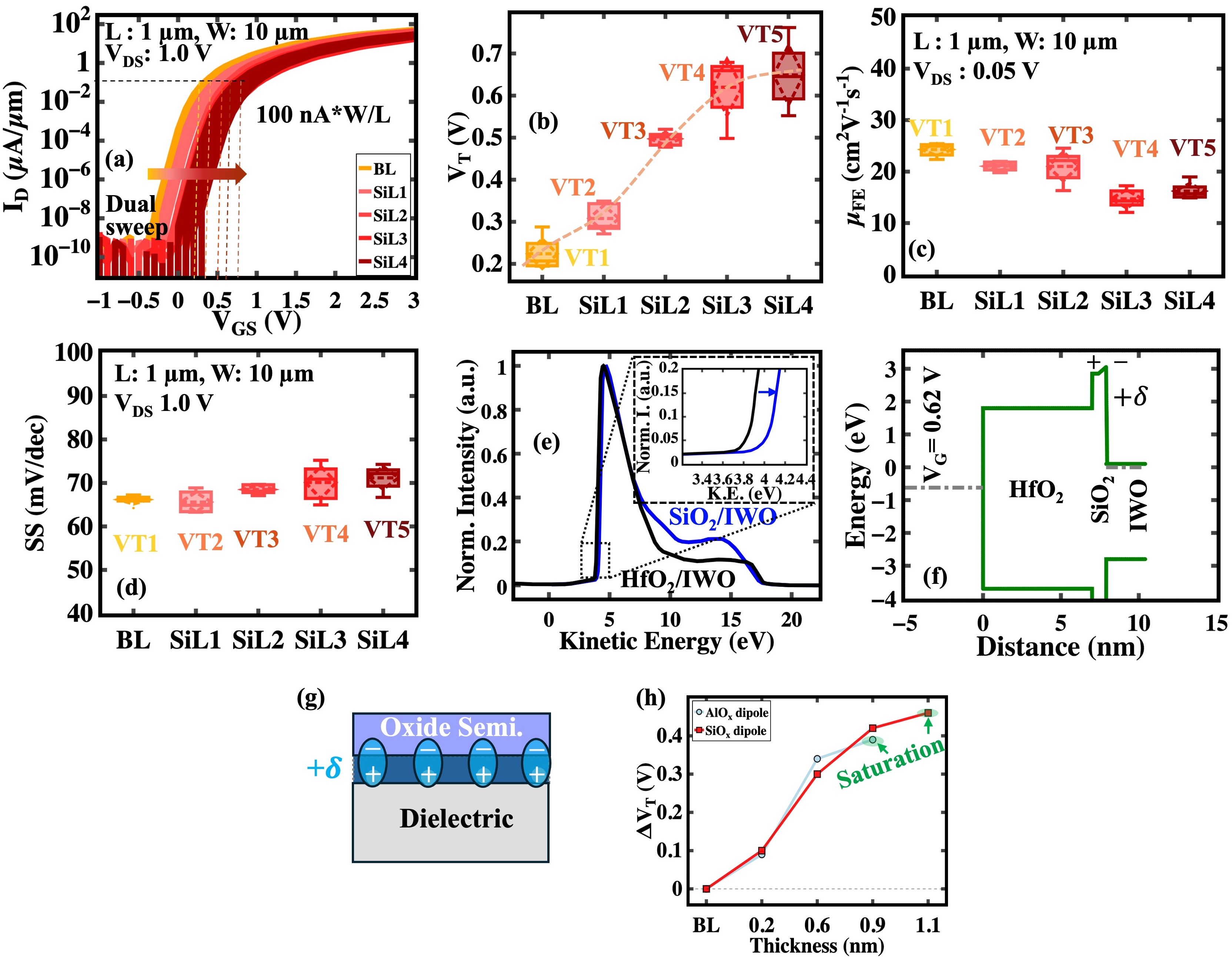}
  \caption{(a) Transfer characteristics (I\textsubscript{D}-V\textsubscript{GS}) of the BL and SiL device, measured at V\textsubscript{DS} 1 V. Here L = 1 $\upmu$m and W = 10 $\upmu$m. SiL devices clearly exhibit a positive V\textsubscript{T} shift. (b) V\textsubscript{T} distributions for different devices, showing that SiL devices achieve a maximum V\textsubscript{T} increase of $\sim$ 500 mV compared to the BL. Here, V\textsubscript{T} was extracted using constant current method at 
100 nA*Width/Length. (c) $\upmu$\textsubscript{FE} of different devices indicate that $\upmu$\textsubscript{FE} remains largely unaffected by the SiL integration. (d) SS values of BL and SiL devices are within acceptable variation range. (e) UPS spectra acquired under applied bias (-10 V) for HfO\textsubscript{2}/IWO and SiO\textsubscript{2}/IWO samples. Inset shows zoomed view of the secondary electron cut-off region. The SiO\textsubscript{2}/IWO exhibits a higher work-function (4.08 eV) than HfO\textsubscript{2}/IWO (3.87 eV). (f) Simulated energy band-diagram of SiL OSFET obtained using one dimensional (1D) Poisson solver. (g) A positive dipole forms at the SiO\textsubscript{2}/IWO interface. (h) Comparison of $\Delta \text{V}_{\text{T}}$=V\textsubscript{T}-V\textsubscript{BL}, for devices employing SiO\textsubscript{x} and an alternate interfacial dipole material (AlO\textsubscript{x}). Except at a dipole thickness of \(0.6\,\mathrm{nm}\), where AlO\textsubscript{x} shows a slightly larger $\Delta \text{V}_{\text{T}}$, the SiO\textsubscript{x} cases exhibit higher $\Delta \text{V}_{\text{T}}$ as a function of thickness, most noticeably at the saturation point (by \(\sim 50\,\mathrm{mV}\)).
}
  \label{fig:2}
\end{figure}

Figure~\ref{fig:2}(a) shows the transfer characteristics of baseline device (BL) and SiL‑engineered AOSFETs. A clear positive shift in V\textsubscript{T} is observed in the SiL‑integrated devices relative to the BL devices, reaching a maximum increase of $\sim$500\,mV. V\textsubscript{T} was extracted using constant current method at 
100 nA*Width/Length (near subthreshold), to include the effects of conduction via shallow band-tail states, which is characteristic of amorphous OS~\cite{jiang2024positive}. The observed positive shift in V\textsubscript{T} is likely attributed to the formation of a positive interface dipole which drives V\textsubscript{T} shift in the positive direction. We tentatively attribute the observed V\textsubscript{T} shift in SiL devices to an interface dipole. This interpretation is supported by the saturation of V\textsubscript{T} shift beyond a certain dipole layer thickness and our previous extensive investigation using an alternative AlO\textsubscript{x} interfacial dipole layer \cite{athena2025orthogonal}, which showed that AlO\textsubscript{x} induced V\textsubscript{T} shift is dipole‑driven rather than caused by fixed charge, nevertheless further analysis is required. This dipole introduces a built‑in electrostatic potential that raises the OS‑channel energy bands relative to the gate, increasing the gate bias required to turn the device on. Importantly, as shown in Figures~\ref{fig:2}(b–d), integrating the SiL has minimal impact on critical device parameters, such as $\upmu_{\mathrm{FE}}$ and SS, confirming precise V\textsubscript{T} modulation without performance degradation. 

Figure~\ref{fig:2}(e) presents ultraviolet photoelectron spectroscopy (UPS) spectra for \ce{HfO2}/IWO and SiO\textsubscript{x}/IWO stacks, acquired under a negative sample bias (-10 V) to expose the secondary electron cutoff. Details of the UPS measurement is provided in the Supporting Information. The inset shows a zoomed view of the cutoff region used to extract the work function by linear extrapolation of the intensity–kinetic energy edge. The SiO\textsubscript{x}/IWO stack exhibits a higher work function (4.08\,eV) than \ce{HfO2}/IWO (3.87\,eV), aligning with the experimentally observed higher V\textsubscript{T} for the SiL‑engineered devices. Figure~\ref{fig:2}(f) shows the simulated energy‑band diagram obtained from a one‑dimensional Poisson solver for the SiL stacks. Dirichlet boundary conditions are applied at the gate electrode and in the quasi‑neutral semiconductor, with a Neumann (zero‑flux) condition at the far boundary conditions are used only where a metal/semiconductor contact is explicitly included. The interface dipole is implemented as an interfacial potential step equivalently, two interfacial charge sheets of opposite polarity similar to Figure~\ref{fig:2}(f), producing an upward band shift on the OS side of about 0.4 V compared to the baseline case and corroborating the proposed dipole‑induced V\textsubscript{T} modulation in the SiL devices. Details of the calculations are provided in the Supporting Information. Figure~\ref{fig:2}(f) compares the $\Delta \text{V}_{\text{T}}$=V\textsubscript{T}-V\textsubscript{BL}, for devices employing SiO\textsubscript{x} and an alternate interface dipole material (AlO\textsubscript{x}). Except at a dipole thickness of \(0.6\,\mathrm{nm}\), where AlO\textsubscript{x} shows a slightly larger $\Delta \text{V}_{\text{T}}$, the SiO\textsubscript{x} cases exhibit  higher $\Delta \text{V}_{\text{T}}$ as a function of thickness, most noticeably at the saturation point (by \(\sim 50\,\mathrm{mV}\)). This modest difference is possibly due to lower dielectric constant of SiO\textsubscript{x}. The dipole strength is expected to be governed primarily by charge‑neutrality‑level (CNL) alignment at the OS/dielectric interface, because OS materials typically possess higher defect densities (and thus more gap states) than dielectrics, the interfacial CNL is more strongly influenced by the OS than by the dielectric. This interpretation is consistent with our prior experiment\cite{athena2025orthogonal}, in which varying the OS while using an AlO\textsubscript{x} dipole resulted in substantially larger $\Delta \text{V}_{\text{T}}$ shifts compared to the change of dipole layer material in this work. Nevertheless, further theoretical and experimental studies are necessary to clarify the microscopic origin of the interface dipole and its interplay with different dielectrics. However, precise V\textsubscript{T} tuning is achievable through the use of different dipole materials, offering a clear advantages for advanced 3D transistor structures, including gate all around structures.

\begin{figure}
  \centering
  \includegraphics[width=1\textwidth]{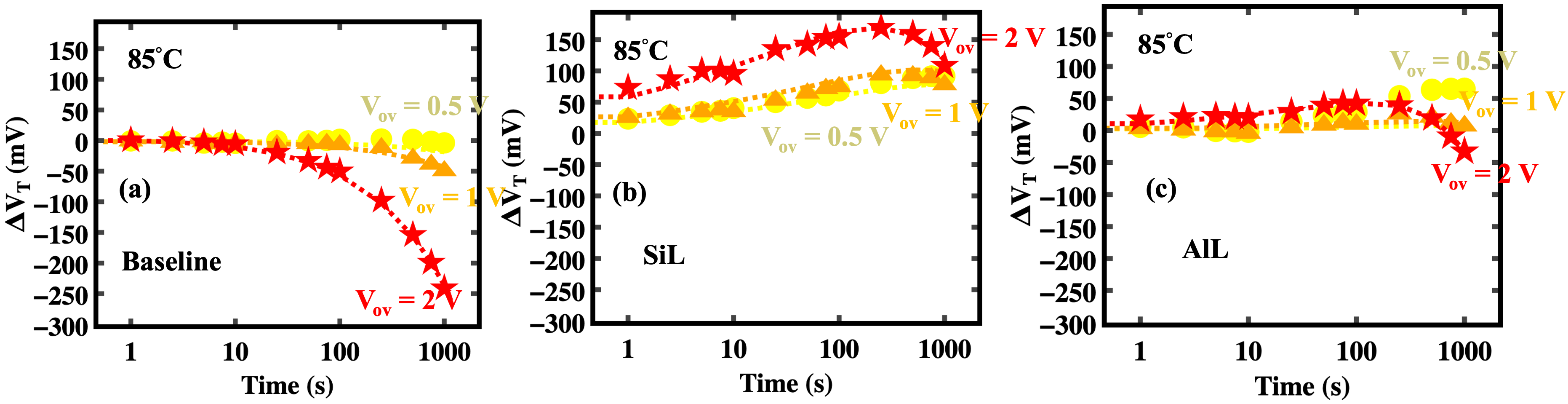}
  \caption{Positive-bias temperature instability of AOSFETs with and without dipole layers, stress applied up to 1000 s at 85 $\degree$C under over-drive biases of 0.5 V, 1.0 V, and 2 V. The dashed lines represent the model fits. (a) BL devices show a negative V\textsubscript{T} shift of $\sim$$-$300 mV at 2 MV/cm, attributed to hydrogen release from the gate dielectric that creates donor-like defects. (b) Dipole-engineered SiL devices exhibit a positive V\textsubscript{T} shift of $\sim$150 mV, indicating that the SiO\textsubscript{x} interfacial layer shifts the dominant degradation mechanism to electron trapping. (c) Devices with an AlO\textsubscript{x} interfacial dipole layer show a smaller negative V\textsubscript{T} shift of around -43 mV at V\textsubscript{ov} of 2 V at 1000 s. Both SiO\textsubscript{x} and AlO\textsubscript{x} interfacial dipole layers eliminate the large negative V\textsubscript{T} shift observed in BL devices enabling improved reliability.
}
  \label{fig:3}
\end{figure}

To evaluate reliability, positive bias temperature instability (PBTI) measurements were conducted at elevated temperatures (85 $\degree$C) under the DC/worst case condition with three different overdrive biases (V\textsubscript{ov} = 0.05 V, 1.0 V, 2.0 V), as shown in Figure~\ref{fig:3} (a,b). BL devices exhibit a significant negative V\textsubscript{T} shift when V\textsubscript{ov} = 2.0 V is applied for 10,000 s, an effect attributed to hydrogen release from the gate dielectric. The released hydrogen (H) can either incorporate interstitially or bind at the oxygen vacancies, where it acts as a shallow donor\cite{aabrar2024reliability,varley2010oxygen,song2019nature}. Each incorporated H contributes an electron to the conduction band, effectively increasing the net donor density in the channel \cite{liu2024unveiling,aabrar2024reliability,yeon2016structural}. Conversely, SiL-integrated devices exhibited a small to moderate positive V\textsubscript{T} shift after stress, indicative of electron trapping rather than H release. Arrhenius-based fits point to a moderately active electron trapping pathway competing with a weaker H release pathway, producing only a modest net shift and improved reliability at the highest V\textsubscript{ov} and long stress times in SiL devices. 

Devices incorporating an alternative AlO\textsubscript{x} dipole layer (AlL) demonstrate a significantly reduced negative V\textsubscript{T} shift at the strongest bias stress voltage (Figure~\ref{fig:3}(c)) that also indicates electron trapping however, to a lesser extent than the SiO\textsubscript{x} case. As mentioned before, the measured data were fitted to an Arrhenius-based reliability model \cite{aabrar2024reliability} to gain deeper insight into the underlying degradation mechanisms, as indicated by the dashed lines in Figure~\ref{fig:3}(a)-(c). Arrhenius-based fits point to a weaker electron-trapping pathway competing with a field-assisted release process, this interplay yields only a small net shift and a much reduced negative shift, at the highest V\textsubscript{ov}. Detailed descriptions of the fitting models and the extracted parameters are provided in the Supporting Information (S6). At room temperature, BL device typically show positive V\textsubscript{T} shift. Stress-induced $\Delta \text{V}_\text{T}$ arises from competition between electron trapping and hydrogen (H)-release--driven electron donation. Arrhenius fitting\cite{jensen1985activation,peleg2012arrhenius} shows that BL devices are dominated by H-release, leading to large negative shifts, while charge trapping is effectively suppressed. For SiL, both trapping and release mechanisms are active and their competition results in a moderate positive $\Delta \text{V}_\text{T}$ at high stress times. AlL shows minimal activity from the positive and negative shift mechanisms, at high $V{ov}$ and long stress times, this device shows a very low magnitude $\Delta \text{V}_\text{T}$ with the sign dependent on V\textsubscript{ov}. Both SiL and AlL layers exhibit smaller V\textsubscript{T} shifts and thus improved PBTI behavior compared to the baseline, however, further investigation is required and will be reported in a sequel to this work.

\begin{figure}
  \centering
  \includegraphics[width=0.85\textwidth]{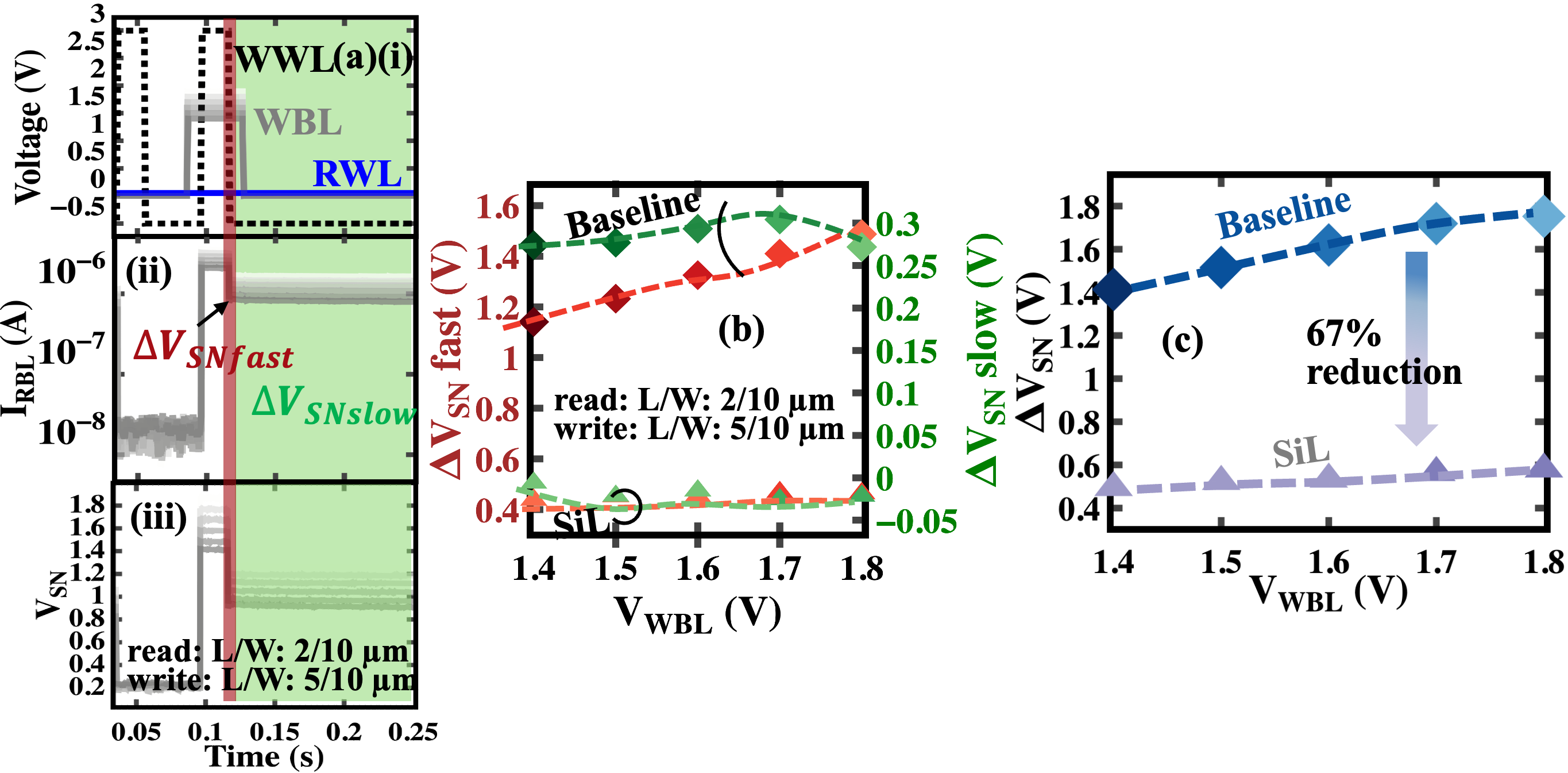}
  \caption{Timing diagram for the BL and SiL GC memory.
  (a)~Measured waveforms as a function of time: (i) WWL, WBL, and RWL voltages, (ii)~ I\textsubscript{RBL}, and (iii)~storage-node voltage V\textsubscript{SN}. 
  V\textsubscript{SN} decays in two stages: an immediate ``fast'' drop coincident with the falling edge of the WWL (red region), followed by a ``slow'' drop driven by post-write leakage (green region). 
  (b)~Distributions of the fast and slow components of V\textsubscript{SN} loss for BL and SiL GCs. SiL GCs show lower V\textsubscript{SN} drop compared to BL GCs.
  (c)~Distribution of the total V\textsubscript{SN} drop. 
  The fast component is governed by the write-transistor V\textsubscript{T}; the higher V\textsubscript{T} achieved with the SiO\textsubscript{x} interfacial layer suppresses mobile charge sharing from the WTR to the SN and WBL, reducing both fast and overall V\textsubscript{SN} drops relative to the BL.
}
  \label{fig:4}
\end{figure}

Figure~\ref{fig:4} illustrates the measured transient characteristics of SiL-engineered GCs and baseline GCs. Figure~\ref{fig:4}(a)(i) shows the applied waveforms for the wordlines and bitlines. To enable multibit data storage, different data voltages V\textsubscript{data} ranging from 1.4 V to 1.8 V are applied at the WBL. The corresponding read bit line current (I\textsubscript{RBL}) is shown in Figure~\ref{fig:4}(a)(ii), while the extracted V\textsubscript{SN} is shown in Figure~\ref{fig:4}(a)(iii). The V\textsubscript{SN} decay occurs in two distinct stages. First, a rapid voltage drop coincident with the falling edge of the WWL pulse (red region), the voltage difference between the applied V\textsubscript{data} and resulting V\textsubscript{SN} after the initial fast drop process is defined as V\textsubscript{SNfast}. 

The rapid V\textsubscript{SNfast} drop is primarily attributed to two mechanisms (1) capacitive coupling between the WTR to the SN\cite{liu2024first}, and (2) mobile charge partitioning from write transistor (WTR) to the WBL and SN. Prior to the WWL falling edge, the WTR channel is in strong accumulation, all trap states in the bandgap are filled and excess carriers accumulate in the conduction band\cite{zheng2024first}. When the WWL voltage drops, the accumulated mobile charge in the WTR channel starts to deplete and flows toward the WBL and SN. Thus, more positive V\textsubscript{T} of the WTR implies reduced charge partitioning. The V\textsubscript{SNfast} is followed by a second stage of gradual decay (V\textsubscript{SNslow}) (green region). Quantitatively, a reduction of $\sim$67\% in the total V\textsubscript{SN} drop is observed for SiL GCs relative to their BL counterparts at WBL data level of 1.8 V. This improvement primarily arises from reduced subthreshold leakage currents and suppressed mobile charge injection enabled by the positive shift in V\textsubscript{T} through SiL dipole engineering.  As a result, the SiL leveraged positive V\textsubscript{T} GCs exhibit a significantly lower overall V\textsubscript{SN} drop compared to BL GCs. 

\begin{figure}
  \centering
  \includegraphics[width=0.75\textwidth]{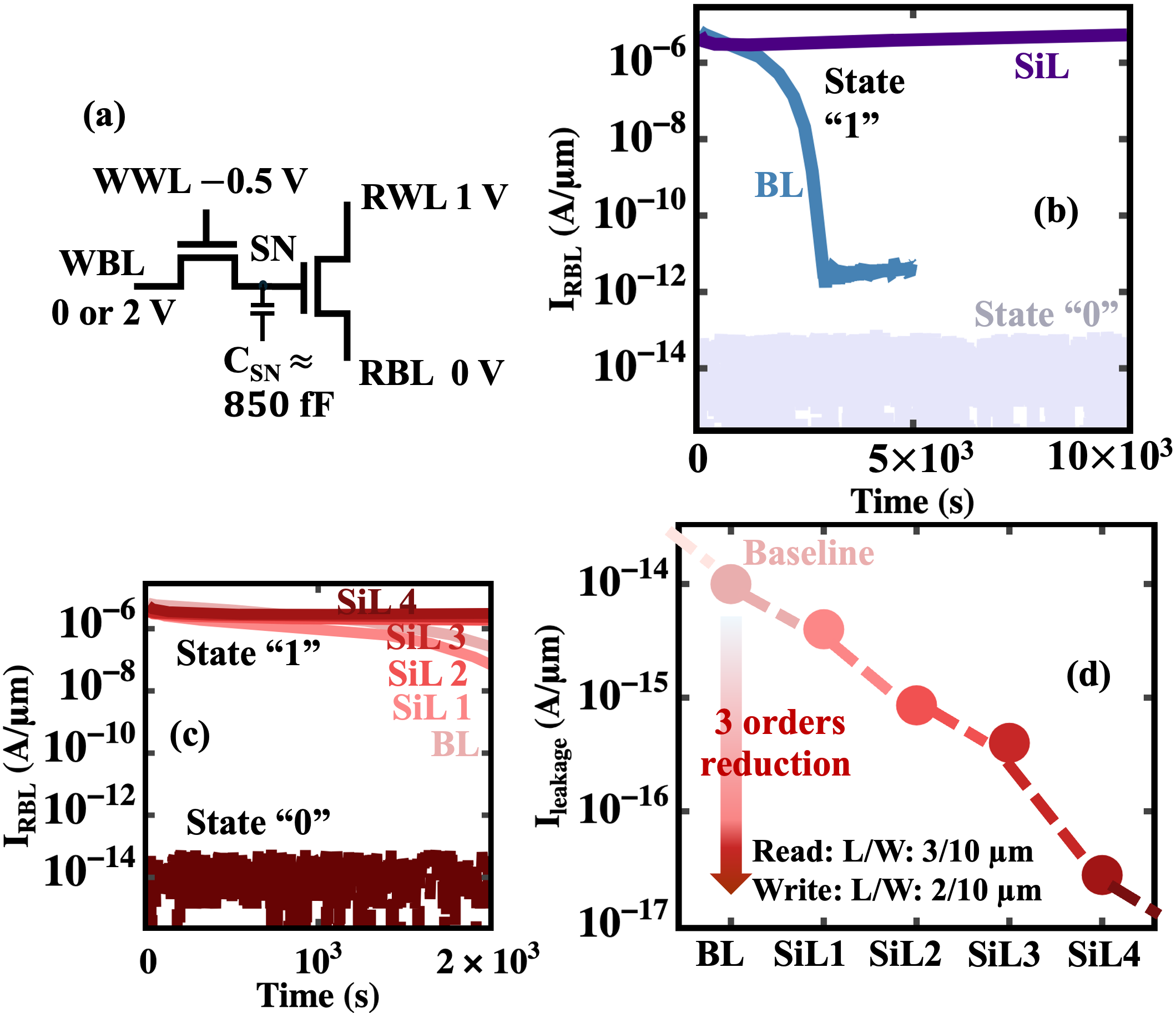}
  \caption{Retention performance of baseline and SiL GCs.
  (a)~Bias condition during standby; WWL is held at $-0.5$\,V. Here, RTR L = 3 $\upmu$m, W = 10 $\upmu$m, WTR L = 2 $\upmu$m, W = 10 $\upmu$m. 
  (b)~The SiO\textsubscript{x}-dipole GC SiL4 shows improved retention (up to 10\,000\,s) compared to BL GC.
  (c)~State-1 retention for the BL and various SiL GCs, measured for 2\,000\,s; State-1 is reported because it is the critical state and typically degrades fastest. 
  (d)~Leakage current extracted from V\textsubscript{SN} degradation (defined as a 0.1\,V drop from the initial value) for BL and SiL GCs. 
  Leakage decreases exponentially with increasing V\textsubscript{T} in the SiL GCs, achieving a three-order-of-magnitude reduction in the SiL4 GCs.
}
  \label{fig:5}
\end{figure}

Figure~\ref{fig:5}(a) details the bias conditions applied during retention measurements. During the standby condition the WWL is held at $-0.5$ V. Figure~\ref{fig:5}(b) demonstrates that SiL GCs achieve significantly improved retention of up to 10,000 seconds, an  improvement over BL GCs which show retention of $<$ 2,500 s. Additionally, retention characteristics for various SiL GCs with different threshold voltages were measured up to 2,000 seconds, as shown in Figure~\ref{fig:5}(c). Leakage currents (storage node leakage) are extracted from retention measurements by measuring the time elapsed for V\textsubscript{SN} to drop by 0.1 V (Figure~\ref{fig:5}(d)) from the initial state. An exponential reduction in leakage current with increasing V\textsubscript{T} is observed in SiL GCs, yielding lower leakage with roughly three orders of magnitude lower than in BL GCs. This directly translates to a significant decrease in refresh-energy demand and an improvement in overall memory efficiency.

\section*{Conclusion}

We demonstrated a volume-less V\textsubscript{T}  tuning method for amorphous AOSFETs by inserting an ultrathin SiO\textsubscript{x} interfacial dipole layer (SiL) in the gate dielectric. The SiL provides a controllable positive V\textsubscript{T}  shift without degrading \(\upmu_{\mathrm{FE}}\) or SS, and it suppresses the negative V\textsubscript{T}  drift under worst-case DC PBTI at \(85\degree\mathrm{C}\). In a 2T GC memory, the SiL stack reduces the $\text{V}_{\text{SN}}$ drop by \(\sim67\%\) versus BL GCs by limiting mobile-charge sharing and extends retention beyond \(10^{4}~\mathrm{s}\), thereby reducing refresh energy.

Relative to our previous work on an AlO\textsubscript{x} dipole-engineered dielectric\cite{athena2025orthogonal}, we investigated SiO\textsubscript{x} to probe how interfacial chemistry (lower \(k\) and different electronegativity) governs dipole-induced V\textsubscript{T}  control and to explore additional candidates for V\textsubscript{T}  tuning in oxide semiconductors. Experimentally, SiO\textsubscript{x} yields a slightly higher V\textsubscript{T}  increase---by \(\sim120~\mathrm{mV}\) compared with AlO\textsubscript{x}---suggesting likely a different effective dipole strength. Whether this magnitude matches simple dipole-strength expectations remains an open question, further investigation is required. Possible next steps include developing a fundamental understanding of the V\textsubscript{T}  change with SiO\textsubscript{x} via modeling and controlled thickness-series experiments, and coupling first-principles calculations with device electrostatics to predict the sign and magnitude of shifts. Establishing these fundamentals for SiO\textsubscript{x} will solidify interfacial-dipole engineering as a rigorous, volume-less V\textsubscript{T}  tuning knob. Because it adjusts V\textsubscript{T}  without altering channel volume, stoichiometry, or doping, this approach is CMOS-compatible and scalable---opening a practical path to back-end compatible dual gate or GAA oxide-semiconductor transistors for high-speed, energy-efficient embedded 3D memory, consistent with the N3XT 3D vision~\cite{radway2021future,hwang2018coming}.


\clearpage

\clearpage

\begin{acknowledgement}

Supported in part by SRC JUMP 2.0 PRISM and CHIMES Center, Stanford Differentiated Access Memory (DAM), SystemX
Alliance, Stanford NMTRI, TSMC-Stanford Joint Development Project (PCM). Part of this work was performed at nano at stanford RRID SCR 026695. The authors gratefully acknowledge Professor Alberto Salleo for helpful discussions and Christina Newcomb for assistance with AFM measurements. F.F.A. would like to thank the support from the Stanford Energy Postdoctoral Fellowship, and Precourt Institute for Energy.

\end{acknowledgement}

\clearpage

\bibliography{achemso-demo}

\end{document}